\documentclass[]{raa}            
\usepackage{graphicx,times}
\usepackage{natbib}

\providecommand{\bm}[1]{\mbox{\boldmath$#1$\unboldmath}}

\begin{document}

   \title{Flares and Magnetic Non-potentiality of NOAA AR 11158
}

   \volnopage{Vol.0 (201x) No.0, 000--000}      
   \setcounter{page}{1}          

   \author{Qiao Song
      \inst{1}
   \and Jun Zhang
      \inst{1}
   \and Shuhong Yang
      \inst{1}
   \and Yang Liu
      \inst{2}
   }

   \institute{Key Laboratory of Solar Activity, National Astronomical Observatories, Chinese Academy of Sciences, Beijing 100012, China {\it qsong@nao.cas.cn, zjun@nao.cas.cn, shuhongyang@nao.cas.cn}\\
        \and
        W. W. Hansen Experimental Physics Laboratory, Stanford University, Stanford, CA 94305-4085, USA {\it yliu@sun.stanford.edu}\\
   }

   \date{Received~~2009 month day; accepted~~2009~~month day}

\abstract{The magnetic non-potentiality is important for understanding flares and other solar activities in active regions (ARs). Five non-potential parameters, i.e., electric current, current helicity, source field, photospheric free energy, and angular shear, are calculated in this work to quantify the non-potentiality of NOAA AR 11158. Benefited from high spatial resolution, high cadence, and continuously temporal coverage of vector magnetograms from the Helioseismic and Magnetic Imager on board the \emph{Solar Dynamics Observatory}, both the long-term evolution of the AR and the rapid change during flares have been studied. We confirmed that, comparing with the magnetic flux, the magnetic non-potentiality has a closer connection with the flare, and the emerging flux regions are important for the magnetic non-potentiality and flares. The main results of this work are as follows. (1) The vortex in the source field directly displays the deflection of horizontal magnetic field. The deflection is corresponding to the fast rotated sunspot with a time delay, which suggests that the sunspot rotation leads to an increase of the non-potentiality. (2) Two areas that have evident changes of the azimuth of the vector magnetic field are found near the magnetic polarity inversion line. The change rates of the azimuth are about $1.3^{\circ}$ h$^{-1}$ and $3.6^{\circ}$ h$^{-1}$, respectively. (3) Rapid and prominent increases are found in the variation of helicity during four flares in their initial brightening regions. The recovery of the increases takes 3-4 h for the two biggest flares (X2.2 and M6.6), while only takes about 2 h for the other two smaller flares (M2.2 and M1.6).
}

   \authorrunning{Song et al.}            
   \titlerunning{Flares and Non-potentiality}  

   \maketitle

%

\section{Introduction}
    \label{S-Introduction}
The solar flare, a common but powerful active phenomenon in the Sun which can release up to the order of 10$^{32}$ erg energy \citep{2005SoPh..230..129K}, has received much attention over the past two centuries. It is generally believed that the enormous energy of the solar flare is provided by magnetic field. When the magnetic field of a solar active region (AR) deviates from a potential (current-free) field configuration to a non-potential configuration, i.e., becomes more sheared and/or twisted, the energy will be stored in the stressed magnetic field and finally it will be released by flares, eruptive filaments, and coronal mass ejections (CMEs) \citep{1987ApJ...314..782G,1989SSRv...51...11S,2008ApJ...675.1637S,2012ApJ...748...77S}. The deviation of an observed vector magnetic field from the potential magnetic configuration is referred by the magnetic non-potentiality.

Non-potential parameters were introduced to measure the magnetic non-potentiality of ARs. The electric current density is directly calculated from the curl of a vector magnetic field. Due to the lack of accurate observation of magnetic field in the higher layers of solar atmosphere, the vertical component of electric current density ($J_{z}$) in the photosphere is widely used \citep{1999FCPh...20..251W,2000JKAS...33...47M}. In a practical calculation, the pixels of a vector magnetogram with low signal-to-noise (S/N) ratios often had unreasonable strong currents. The electric current helicity can avoid this problem by improving the effect of magnetic field strength to the current. As a part of a scalar, the helicity of vertical current ($h_{c}$) is also called the fractional current helicity \citep{1996SoPh..168...75A,1998ApJ...496L..43B}. The angular shear and other two non-potential parameters come from the differences between the observed magnetic field and the potential magnetic field which is extrapolated from observed photospheric field with the potential filed assumption. \citet{1984SoPh...91..115H} defined the angular magnetic shear as the degree of angle between the observed horizontal field and the horizontal component of potential field. \citet{1993SoPh..148..119L} pointed out that the shear angle which is defined by the angle between the vector of observed magnetic field and the vector of computed potential field is a better indicator for the twisting effect of the vector magnetic field. \citet{1996ApJ...456..861W} assigned a sign to the shear angle to introduce the angular shear ($\theta_{as}$). The sign is determined by the direction of rotation from the horizontal component of potential field vector to the horizontal component of the observed magnetic field vector with a clockwise rotation considered negative. Since the angular shear only needs the angles of vectors, it refrains from errors of the calibration of magnetic field strength. \citet{1981SoPh...73..257H} defined the source field ($\vec{B}_{s}$) as the vector departure while the angular shear as the angular departure. The source field is a vectorial and comprehensive expression of magnetic non-potentiality and it is given by $\vec{B}_{s}=\vec{B}_{obs}-\vec{B}_{pot}$, where the subscripts $obs$ and $pot$ refer to the observed and potential magnetic field. The photospheric free energy density ($\rho_{free}$) is proportional to the square of the strength of the source field.

Through the parameters, the relationship between flares and the non-potentiality has been studied. \citet{2005ApJ...628..501S} sampled 95 ARs and concluded that flares are 2.4 times more frequently in non-potential ARs than in potential ARs. Some researches about the spatial relationship showed that initial bright kernels of flares are located in vicinities of peaks of some non-potential parameters \citep{1984SoPh...91..115H,1994SoPh..155...99W}. An increase of magnetic shear was found after several X-class flares \citep{1994ApJ...424..436W,2002ApJ...576..497W}. \citet{2001SoPh..204...11D} identified that the magnetic shear changed its sign in the filament channel of NOAA AR 9077. Rapid and irreversible increases in the horizontal magnetic field ($B_{h}$) during flares were also found \citep{2009ApJ...690..862W,2012ApJ...745L..17W,2012ApJ...745L...4L,2012ApJ...748...77S} in the areas around the magnetic polarity inversion line (PIL). Furthermore, it has been shown that the emergence of magnetic flux plays an important role in the development of the non-potentiality and eventually produces flares/CMEs. New emerging bipoles were found cospatial with significant vertical electric currents \citep{1994SoPh..155...99W,1996ApJ...462..547L}. \citet{2004ApJ...615.1021W} further noted that some emerging flux regions (EFRs) brought an opposite sign to the dominant helicity and the flare/CME initiation site which was characterized by the opposite sign and magnetic flux cancelation. Other studies also showed the emergence of twisted magnetic flux ropes and the injection of opposite helicity involved with strong X-class flares \citep{2007ApJ...657..577D,2010ApJ...720.1102P}.

It is well known that vector magnetic field observation is irreplaceable for the study of the non-potentiality. The Helioseismic and Magnetic Imager (HMI; \citealt{2012SoPh..275..229S}) on board the \emph{Solar Dynamics Observatory} (\emph{SDO}; \citealt{2012SoPh..275....3P}) provides an uninterrupted high cadence and high spatial resolution full-disk vector magnetic field observation without atmosphere effect. This unprecedented observation allows us to investigate the evolution of an AR in a long-term period and the short-term change of the non-potentiality during flares of the AR simultaneously. In order to achieve this goal, we focus on temporal and spatial variations of the photosphere magnetic field in detail. We chose NOAA AR 11158 as the target for its complex magnetic structure, rapid evolution, and abundant activities. The AR erupted the first X-class flare of Solar Cycle 24, therefore plenty of observational studies have paid attention to it \citep{2011ApJ...738..167S,2012ApJ...744..166T,2012ApJ...747..134M,2012AN....333..125B,2012ApJ...749...85G,2012ApJ...749L..16D,2012SoPh..tmp..131A,2012SoPh..tmp..183P,2012ApJ...752L...9J}.

The data and the evolution of non-potentiality of the AR are briefed in Section~\ref{S-Observation}. The confirmation of earlier results is presented in Section~\ref{S-Confirmation}. Section~\ref{S-Result} reports new results of the changes and evolutions of the non-potentiality during four major flares of the AR. Section~\ref{S-D&S} summarizes the results and discusses the possible reasons of the changes and evolutions.

\section{Observation}
   \label{S-Observation}
The vector magnetograms we used were observed by \emph{SDO}/HMI from 2011 February 12 to 16 with a spatial resolution of $\sim1''$ and a cadence of 12-min and then reduced by the HMI science data processing pipeline\footnote{http://jsoc.stanford.edu/jsocwiki/ClickHereForDataRelease}. The Stokes parameters $I$, $Q$, $U$, $V$ from the Fe I 6173 \AA$ $ spectral line were calculated by Very Fast Inversion of the Stokes Vector (VFISV) algorithm \citep{2011SoPh..273..267B}. Afterwards, the `minimum energy' \citep{1994SoPh..155..235M,2006SoPh..237..267M,2009SoPh..260...83L} solution was used to resolve the 180-degree ambiguity and then the images were remapped with a Lambert cylindrical equal area projection \citep{Hoeksema}. In the end, the magnetic field vector in each pixel of the data set was turned to a three-component in a three-dimensional rectangular coordinate system with a vertical component $z$ and two horizontal components $x$ and $y$.

It is clear that the accuracy of calculated non-potential parameters depends on the reliability of vector magnetograms. For the strong magnetic field of AR 11158 we could use a high threshold to filter noises. In this work, only the pixels with the strength of vector magnetic field ($\vec{B}$) stronger than 300 G were considered. Because the work is about the variation rather than the absolute value, the threshold would only cause a minor effect. The simultaneous ultraviolet 1600 {\AA} images taken by the Atmospheric Imaging Assembly (AIA; \citealt{2012SoPh..275...17L}) on \emph{SDO} were also employed to investigate the flare ribbons.

AR 11158 was a complex quadrupole system that erupted one X-class, 5 M-class, and 56 C-class flares from 2011 February 9 to 21. The data set in our work covered the entire quick growth phase and 4 major flares ($\geq$M1.0, see Table 1). In the growth phase, numerous magnetic flux emerged and 10 key EFRs were identified in the vector magnetograms with a $205''\times 125''$ field of view (FOV) as shown in Figure 1. The EFRs were identified by the emerging time, the motion, the magnetic flux of the opposite polarities and the appearance of arches on AIA extreme ultraviolet images.

\begin{table}
\caption{Four major flares ($\geq$M1.0) of NOAA AR 11158 from 2011 February 12 to 16.}
\label{T-1}
\begin{center}
\begin{tabular}{ccccc}
\hline\noalign{\smallskip}
Date & Start & Peak & End  & X-ray \\
     & time  & time & time & class\\
\hline\noalign{\smallskip}
Feb 13 & 17:28 & 17:38 & 17:47 & M6.6 \\
Feb 14 & 17:20 & 17:26 & 17:32 & M2.2 \\
Feb 15 & 01:44 & 01:56 & 02:06 & X2.2 \\
Feb 16 & 14:19 & 14:25 & 14:29 & M1.6 \\
\noalign{\smallskip}\hline
\end{tabular}
\end{center}
\end{table}

\begin{figure}\centering\includegraphics[width=0.95\textwidth,clip]{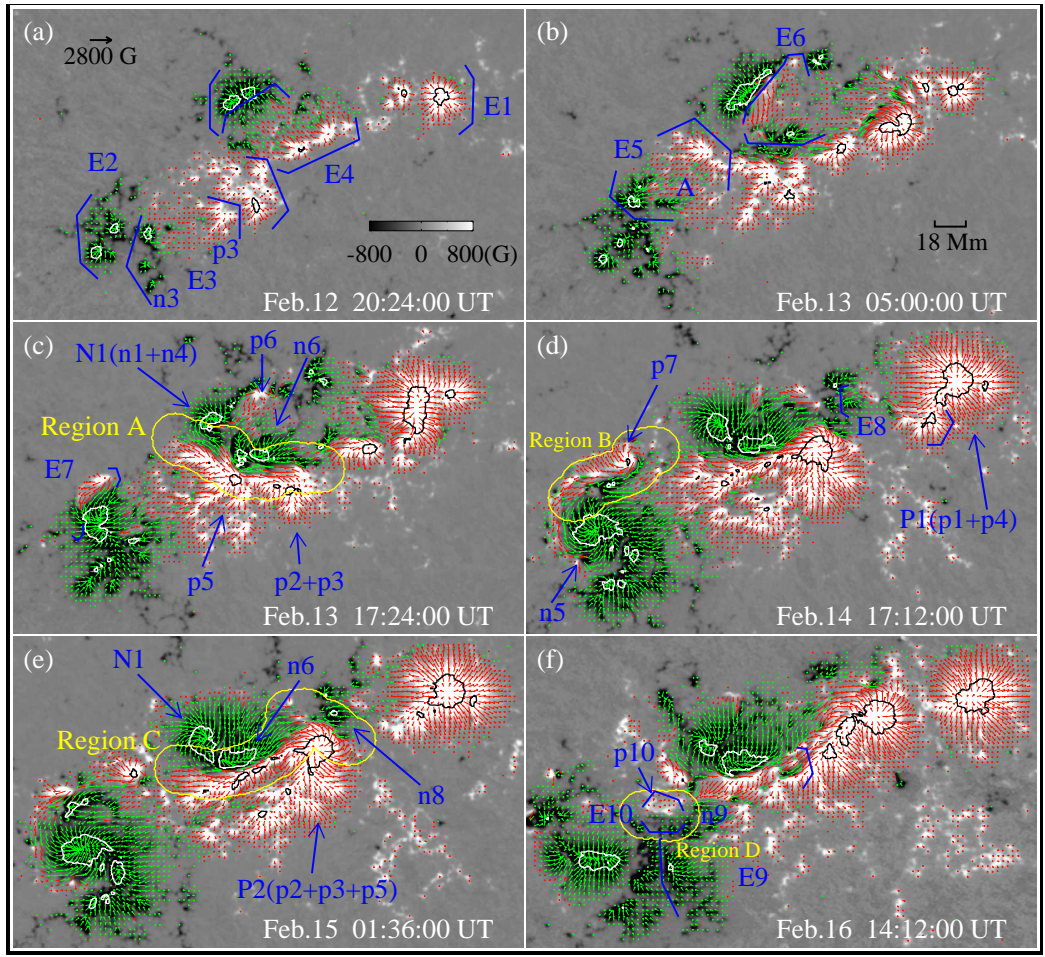}
\caption{Overview of vector magnetograms in NOAA AR 11158 from 2011 February 12 to 16. The vertical magnetic component is presented by black-white patches and isogauss contours with levels 1500 G (black) and $-$1500 G (white). Green and red arrows denote the horizontal magnetic component. The field of view (FOV) in each panel is about $205''\times 125''$. E1--E6 in (a) and (b) indicate six major emerging flux regions (EFRs) and E7--E10 in (c), (d) and (f) show other four smaller ones. Letters `p' and `n' with numbers indicate the positive and negative polarities that come from corresponding EFRs, respectively. (c)--(f) show the vector magnetic fields just before four major flares (M6.6, M2.2, X2.2 and M1.6) onset times, respectively. The yellow contours on the four panels outline regions A, B, C and D (see Section~\ref{S-Result}).}\label{fig:1}
\end{figure}

The first EFR (E1) started emerging when the AR was near the east limb of the Sun at about 23:00 UT on February 9. The second one (E2) followed it and came out from the southeast of E1 (see Figure 1a). E3 and E4 appeared one after another in the middle of February 12 and contributed $\sim 1 \times 10^{21}$ Mx and $\sim 3 \times 10^{21}$ Mx total unsigned flux to the AR, respectively. The positive and the negative polarities of E3 (p3 and n3, similarly hereinafter) separated at an average speed of $\sim$ 0.7 km s$^{-1}$. Finally, they merged with the two polarities of E2.

The crucial E5 and E6 sprang up almost simultaneously and doubled the unsigned flux of the AR on February 13. Meanwhile, a $\delta$-sunspot that contained p5, n6 and their predecessors was formed. As can be seen in Figure 1b, the vector magnetic field is obviously horizontal in area A. It indicates that E5 was emerging in the area with a similar direction of E2 and E3, and brought $\sim 5 \times 10^{21}$ Mx flux to the AR at 17:24 UT on February 13. The two polarities of E5 separated at a speed of $\sim$ 1.5 km s$^{-1}$. We noted that the emergence of E6 had a similar location but a contrary direction with the previous E4. The contrary direction led to a magnetic cancelation between p6 and N1 (the combination of n1 and n4) and eventually caused the vanishing of p6. Furthermore, n6 met with the fast moving p5 and soon the first major flare of the AR erupted (Figure 1c).

E7 was a smaller but important EFR that came out from a location close to n5. On February 14, when p7 was accelerated and met with negative polarities from area A, the second major flare of the AR broke out (Figure 1d). At the northwest of the AR, another small EFR (E8) emerged from an area near P1 (the combination of p1 and p4), simultaneously. The n8 encountered P2 (the combination of p2, p3 and p5) on February 15 and it might trigger the biggest flare of the AR, i.e., the first X-class flare of Solar Cycle 24. The last major flare of our data set erupted on February 16, when E9 and E10 were emerging (see Figure 1f). The cancelation between p10 and n9 might cause this flare.

\section{Confirmation of earlier results}
   \label{S-Confirmation}
Through the excellent HMI vector magnetograms, we confirmed many earlier results of the non-potentiality in the complex AR 11158.

As mentioned above, the collision and cancelation between the opposite polarities of the different EFRs play an important role to the occurrence of the flares of the AR, which is in line with a recent study about another complex AR \citep{2012AJ....143...56Y}. Furthermore, as former studies shown \citep{1994SoPh..155...99W,1996ApJ...462..547L}, EFRs brought significant vertical electric currents and other non-potential parameters to the AR. The first five EFRs (E1-E5) emerged with approximately equal positive and negative current helicity. When E6 rapidly sprang up in the opposite direction to its predecessor, the balance was disrupted. It is noted that N1 changed its sign from negative to positive (Figures 2a and 2b). After the emergence of E5 and E6, N1, n6 and p5 concentred to three new current cores with different signs (Figure 2c). As we can see in Figure 2d, a strong negative shear belt ($\leq -80^{\circ}$) outlined the PIL. The increase of this belt during the flare was also found in our work as well as previous works which used slight different definitions of angle \citep{1994ApJ...424..436W,2002ApJ...576..497W,2012ApJ...748...77S}. The box in Figure 2d indicated an area with opposite angular shear that will be discussed in Section~\ref{S-D&S}.

\begin{figure}\centering\includegraphics[width=1\textwidth,clip]{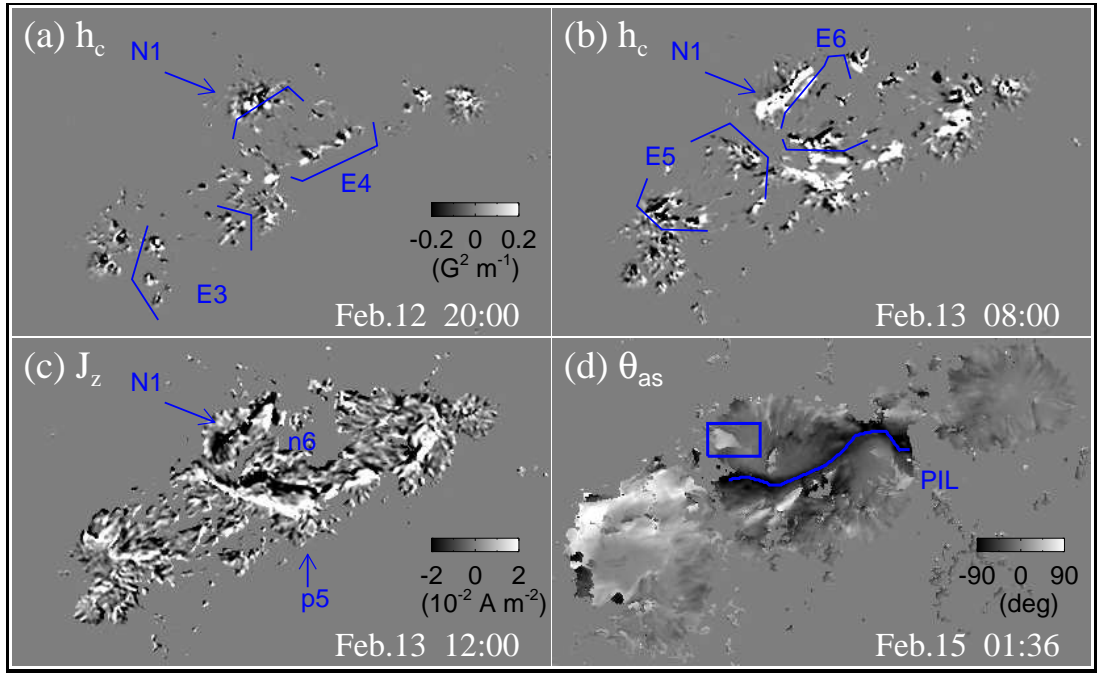}
\caption{Maps of fractional current helicity ($h_{c}$), vertical current ($J_{z}$) and angular shear ($\theta_{as}$). The FOV and EFRs are as same as Figure ~\ref{fig:1}. The curve in (d) shows the magnetic polarity inversion line (PIL) and the box indicates an opposite sign area of $\theta_{as}$ which is discussed in Section~\ref{S-D&S}.}\label{fig:2}
\end{figure}

As illustrated by Figure 3, the evolutions of the magnetic flux and non-potential parameters were not uniform. \citet{2008ApJ...686.1397P} distinguished the magnetic helicity accumulation before a flare into two phases, a monotonically increasing phase and a relatively constant phase. The two phases (I and II) that before the X-class flare were confirmed in the evolution of AR 11158 and the AR turned to an undulate decrease phase (III) after the flare (Figure 3b). As listed in Table 2, the growth of the AR was mainly at phases I and II. The unsigned flux, current and helicity had a fast and a slow increase in phases I and II, respectively. However, both the unsigned angular shear and total free energy had a sharp rising in the middle of phase I and continued a fast increase in phase II. The variations of the current, helicity, and free energy were declined in phase III, while the variations of the flux and angular shear were still rising.
The major flares happened only when the flux and non-potentiality of the AR reached a high level. During and after phase III, the occurrence frequency and intensity of the major flare decreased, while the flux was still relatively high. It confirms that the non-potentiality, instead of the magnetic flux, has a stronger connection with the flare.
The total positive and the negative fluxes kept balance during phase II while the positive and negative currents had a minor imbalance through all the three phases. The negative helicity peaked when the M2.2 flare broke out and dropped immediately after the X2.2 flare erupted. On average, the negative shear was as $\sim 2.3 $ times as the positive shear during the five days.

\begin{figure}\centering\includegraphics[width=0.7\textwidth,clip]{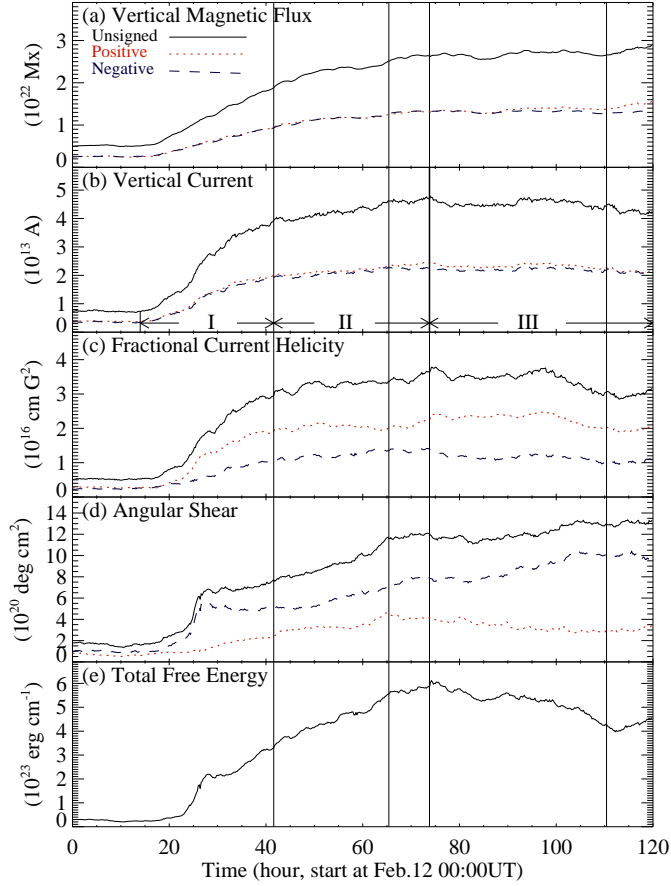}
\caption{Evolution of magnetic flux and four non-potential parameters of AR 11158 in five days. (a) is the variation of total unsigned, positive, and negative magnetic flux. The four long vertical lines in each panel indicate the onset times of the four flares. (b)--(d) have the same line styles as (a) but for the variations of vertical currents, current helicities, and angular shears. (e) is the variation of total free energy. (b) also give the three phases of the evolution (see Table~\ref{T-2}).}\label{fig:3}
\end{figure}

\begin{table}
\caption{Three phases of evolution of the non-potentiality of AR11158 from 2011 February 12 to 16. The first line of each phase is the average rate of daily change and the second line gives the percent of the change in corresponding phase.}
\label{T-2}
\begin{center}
\begin{tabular}{ccccccc}
\hline\noalign{\smallskip}
 & Start Time     & Flux   & Current    & Helicity    & Angular Shear   & Free Energy  \\
Phase  & End Time & (10$^{22}$ Mx & (10$^{13}$A & (10$^{16}$cm G$^{2}$ & (10$^{20}$deg cm$^2$ & (10$^{23}$erg cm$^{-1}$ \\
      & (UT)  &  day$^{-1}$) & day$^{-1}$) & day$^{-1}$) & day$^{-1}$)  & day$^{-1}$) \\
\hline\noalign{\smallskip}
I     & Feb 12 14:00 &1.03 & 2.78 & 2.10 & 5.34 & 2.67\\
  & Feb 13 17:28     &  (47.8\%) & (66.8\%)  & (64.0\%)  & (46.0\%) & (50.3\%) \\
\hline\noalign{\smallskip}
II     & Feb 13 17:28  &  0.560 & 0.559 & 0.503 & 3.25 & 1.90\\
  & Feb 15 01:44  & (26.1\%) & (15.7\%)  & (17.8\%)  & (32.6\%) & (41.6\%) \\
\hline\noalign{\smallskip}
III     & Feb 15 01:44  & 0.128 & -0.226 & -0.241 & 0.658 & -0.673\\
 & Feb 16 23:48 & (8.53\%) & (-9.07\%) & (-12.2\%) & (9.44\%) & (-21.1\%)\\
\noalign{\smallskip}\hline
\end{tabular}
\end{center}
\end{table}

 Figure 4 shows difference maps of the $B_{z}$, $B_{h}$, $h_{c}$ and $\rho_{free}$ for the X2.2 flare. The flare ribbons appeared a double-J configuration laid astride the PIL. Besides of smaller enhanced and weakened areas, two bigger areas exit the difference map of $B_{z}$. Area B was due to the emergence of magnetic flux, while area C came from the fast moving P2 (Figure 4a). The fast changes of $B_{h}$ around PIL during the flare are confirmed in the present work as well as previous works (Figure 4b). The $B_{h}$ of the X2.2 flare had significant enhancement in the junction of these two J-shape structures where \citet{2012ApJ...745L..17W} found ~30\% increase during the flare. The $B_{h}$ of all the four flares appeared obvious enhancement on the flare ribbons and reduced in the outskirts which is consistent with a previous result \citep{2009ApJ...690..862W}. The biggest enhanced areas in $h_{c}$ was in the same location with the negative polarity n6. The greatest receded areas in this parameter was cospatial with the negative polarity N1 (Figure 4c). The $\rho_{free}$ also had significant enhancement in the junction of these two J (Figure 4d).

\begin{figure}\centering\includegraphics[width=0.9\textwidth,clip]{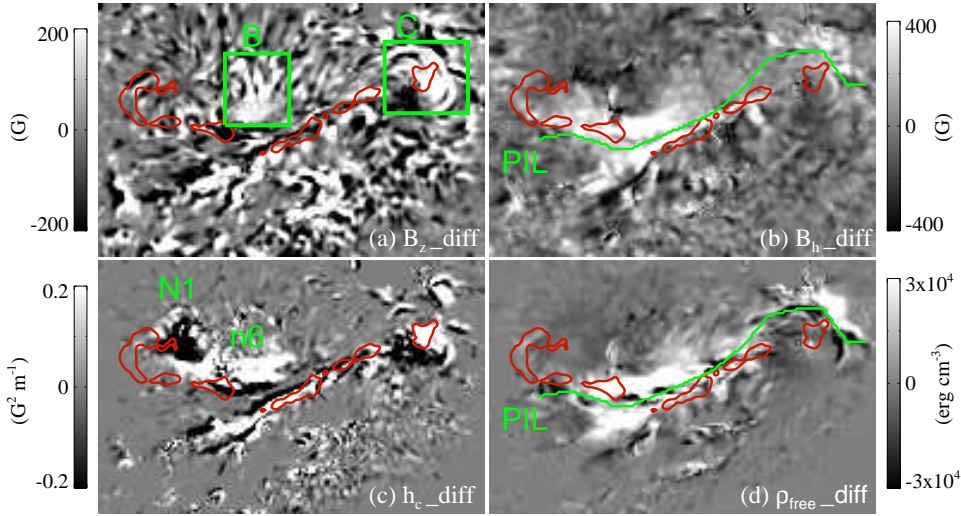}
\caption{Different maps of vertical magnetic component ($B_{z}$), horizontal magnetic component ($B_{h}$), $h_{c}$ and free energy density ($\rho_{free}$) for the X2.2 flare at 01:36 and 02:24 UT on February 15. The FOV of each panel is about 80$''$$\times$53$''$. The red contours denote flare ribbons at 01:47 UT and the green curves show the PIL. The boxes in (a) indicate enhanced areas of $B_{z}$.}\label{fig:4}
\end{figure}

\section{New results}
   \label{S-Result}

The $\vec{B}_{s}$ and $B_{z}$ of P2 are mapped in Figure 5 to give an example of vortexes of the $\vec{B}_{s}$. The $\vec{B}_{s}$ is represented by arrows with length proportional to the relative field strength and the $B_{z}$ is presented by gray-scale patches. The FOV of each panel is about $25'' \times 25''$. The blue and yellow boxes in Figure 5b are as same as those in Figure~\ref{fig:6} to show the locations of two small areas that will be described below. By definition, the strength of $\vec{B}_{s}$ is proportional to the square root of $\rho_{free}$. Therefore, its distribution and variation are similar to the free energy. However, the direction of $\vec{B}_{s}$ has additional information about the non-potentiality. Around the PIL of the AR, the strength of $\vec{B}_{s}$ was stronger and the direction was from negative to positive magnetic polarities (Figure 5a). Then an anticlockwise vortex appeared on the west end of the flare ribbons about 15 h before the X2.2 flare (Figure 5b). As Figure 5c shown, the vortex was still evident one day after the flare when P2 was rotating around its center and canceling with n8. The vortex disappeared on the next day as well as n8 (Figure 5d). In the meantime, the unsigned current helicity decrease $\sim 4.1 \times 10^{15}$ cm G$^{2}$ (34\%) in the area.

The $\vec{B}_{s}$ of N1 and n5 also had vortexes. The $\vec{B}_{s}$ of N1 formatted a clockwise vortex about 9 h before the first major flare and the vortex developed in the next two days. The east end of ribbons of the M6.6 flare overlapped with this vortex, while ribbons of the X2.2 flare were partly overlapped. An anticlockwise vortex that around n5 formatted in late February 13, then enhanced about 13 h before M2.2 flare, and finally vanished on February 16. Simultaneously, n5 rotated at least $180^{\circ}$ about its center.

\begin{figure}\centering\includegraphics[width=0.55\textwidth,clip]{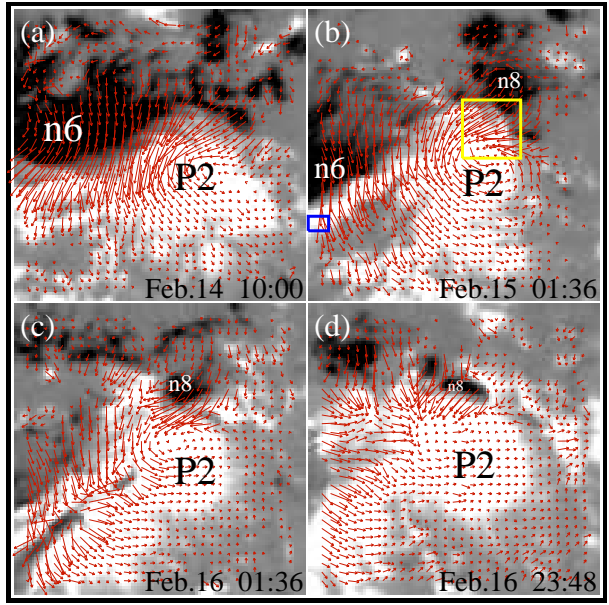}
\caption{Vortex of source field ($\vec{B}_{s}$) of P2. The $\vec{B}_{s}$ is represented by arrows with length proportional to the relative field strength. The $B_{z}$ is presented by gray-scale patches. The FOV of each panel is about $25'' \times 25''$. The blue and yellow boxes in (b) are as same as those in Figure~\ref{fig:6} to show the locations.}\label{fig:5}
\end{figure}

Figure 6 shows two areas that had evident changes of the direction of $B_{h}$ (e.g. azimuth). Green and red arrows denote the $B_{h}$. The $B_{z}$ is presented by black-white patches and isogauss contours with levels 1500 G (black) and $-$1500 G (white). The FOV of each panel is about $12.5'' \times 12.5''$. The blue arrows indicate the average azimuths in the boxes. The northwest part of P2 started a directly interaction with the opposite polarity n8 before the X-class flare (see Figure 6a and the yellow box in Figure 5b). After the flare, the two polarities canceled with each other even faster than preflare. As illustrated by the arrows, the average azimuth inside the yellow box anticlockwise rotated $\sim 7.5^{\circ}$ and turned to more parallel to the PIL. The cancelation continued even 10 h after the flare (Figure 6b). About $\sim 3.9 \times 10^{20}$ Mx negative magnetic flux was canceled in the area from 00:00 UT on February 15 to 17:00 UT on February 16. Almost all negative magnetic flux in the area (96\%) was canceled. Meanwhile, the rotation of $B_{h}$ turned to clockwise which means the vector magnetic field of P2 became more potential. After n8 disappeared by the cancelation, the average azimuth had a change of $54^{\circ}$ to that of preflare. The change rate was about $1.3^{\circ}$ h$^{-1}$.

In the southeast neighborhood of P2, a complex area also had an evident change of the average azimuth (see Figure 6d and the blue box in Figure 5b). A strong PIL was across the area and the negative polarities of E9 moved pass the area (also see Figure 1f and Figure 2d). In the meantime, several positive polarities merged in the area. the change of average azimuth was remarkable in the blue box between two positive polarities that were merging (p11 and p12). The change of average azimuth inside the box during 8 h was about $29^{\circ}$ with a rate of $3.6^{\circ}$ h$^{-1}$ (Figure 6f).

\begin{figure}\centering\includegraphics[width=1\textwidth,clip]{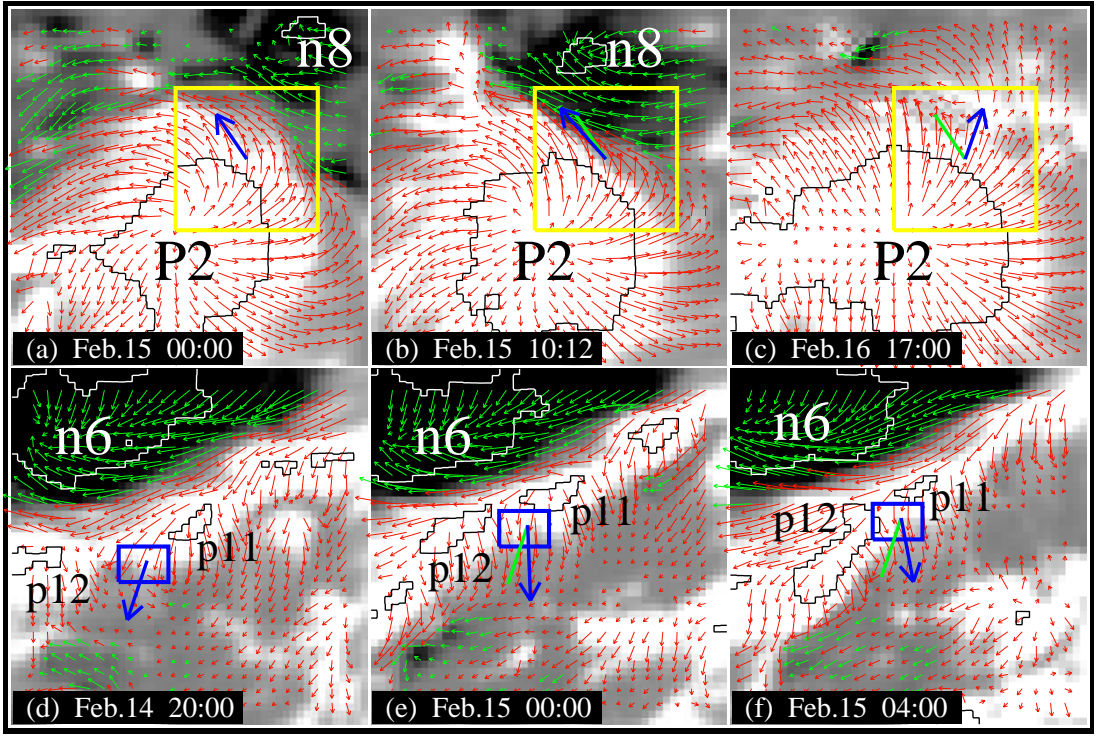}
\caption{Changes of the direction of $B_{h}$ (e.g. azimuth) in two small areas. The FOV of each panel is about $12.5'' \times 12.5''$. The set of vector magnetograms is similar to that in Figure ~\ref{fig:1}. The blue and yellow boxes are as same as those in Figure~\ref{fig:5}. The blue arrows indicate the average azimuths in the boxes.}\label{fig:6}
\end{figure}

 Four regions are outlined by the yellow contours on panel (c)-(f) in Figure~\ref{fig:1} to investigate the change of the non-potentiality for the four flare events. Each pixel of the regions has the distance less than 15 Mm to the PIL \citep{2007ApJ...655L.117S}. The main part of initial brightening in 1600 {\AA} of the flare ribbons are located in the regions. These four selected regions are assigned as regions A, B, C, and D, respectively. The variations of $B_{z}$ and the four non-potential parameters during the processes of the four flares in these initial brightening regions are detailed in Figure 7. Abscissas are times that start at Feb. 13 10:36, Feb. 14 10:12, Feb. 14 18:36 and Feb. 16 07:36 UT, respectively. The gray shadow on each panel indicates the duration of the corresponding flare.

 Most of the parameters increase before the flares and decrease after them. We found that the unsigned magnetic flux in regions A and C continued increasing and reached their peaks 2-3 h after the flares, while the other two did not show this conspicuous change. These variations of regions A and C are more significant if we enlarge the regions to include the entire P2 and N1. The unsigned current helicity had a fast and significant increase in each region and then slowly reverted. The dot line in panel b3 of Figure 7 indicates the time when p7 began its fast proper motion and about one hour later the unsigned helicity started a rapid increase. The unsigned current had similar but temperate increases except in region D. The unsigned angular shear rose before the first two flares and started declining immediately before the onset of these flares. The total free energy in region C had a rapid increase more than 10\% during the X2.2 flare while other regions had smaller increases. The reason of these changes and other issues will be discussed in the next section.

\begin{figure}\centering\includegraphics[width=0.7\textwidth,clip,bb=35 188 515 640]{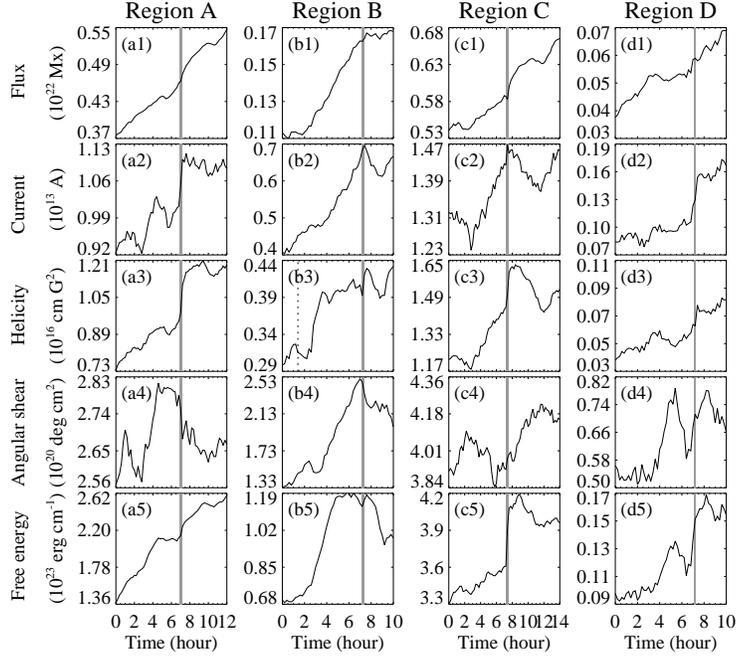}
\caption{Variations of the vertical magnetic flux and the four non-potential parameters through processes of the flares. \emph{Rows}: Unsigned vertical magnetic flux, unsigned vertical current, unsigned fractional current helicity, unsigned angular shear, and total free energy; \emph{Columns}: four regions for the flares as shown in Figure \ref{fig:1}. Abscissas are times that start at Feb. 13 10:36, Feb. 14 10:12, Feb. 14 18:36 and Feb. 16 07:36 UT, respectively. The gray shadow on each panel indicates the duration of the corresponding flare. The dot line in (b3) shows the time when the moving of p7 and the rotating of n5 are both accelerated.}\label{fig:7}
\end{figure}

\section{Discussion and Summary}
   \label{S-D&S}
Benefited from the state-of-the-art vector magnetograms from $SDO$/HMI, we studied long-term evolution of AR 11158 and short-term change of the non-potentiality during four major flares in the AR. We confirm that the non-potentiality, instead of the magnetic flux, has a closer connection with the flare. The EFRs were important for the non-potentiality and flares of the AR because they increased the non-potentiality and brought the crucial opposite sign of non-potentiality to the AR. AR 11158 was formed by contributions from 10 key EFRs that simultaneously occurred or quickly succeeded within five days. The interaction between opposite polarities of the EFRs may be a likely cause of some major flares. Some crucial EFRs could impact on the imbalances of the whole AR. The evolution of non-potential parameters were fast increasing as major EFRs were emerging in phase I. The major flares happened only when the flux and non-potentiality of the AR reached a high level. The current, helicity, and free energy declined in phase III, while the flux still rose. The rapid increase of $\rho_{free}$ and $h_{c}$ around PIL during the flare was found in our work as that in previous. We also identified decrease at outskirts of flare initial brightening regions for the two non-potential parameters as well as earlier studies.

The sunspot n5 had fast and significant rotation before the M2.2 flare and \citet{2012ApJ...744...50J} had found that the sunspot P2 had a rapid rotation before the X2.2 flare. The relationship between the sunspot rotation and the major flare had been discussed in previous works \citep{2007ApJ...662L..35Z,2008MNRAS.391.1887Y}. The $\vec{B}_{s}$ vortex in our work directly displayed the deflection of $B_{h}$. The evolution of the vortexes had a few hours of delay to the rapid rotations of the sunspots. It might suggest that the sunspot rotation leads to an increase of the non-potentiality of the AR.

\citet{2012ApJ...745L...4L} used the nonlinear force-free field (NLFFF) to reconstruct three-dimensional coronal magnetic field and explained the increase of $B_{h}$ with the tether-cutting model \citep{2001ApJ...552..833M} for the M6.6 flare event of AR 11158. \citet{2012ApJ...748...77S} gave a similar result for the X2.2 flare of the AR. According to the model and the NLFFF reconstruction, the coronal magnetic field collapses after the flare. Vector magnetic field of photosphere becomes more inclined, i.e., $B_{h}$ increases and $B_{z}$ decreases. However, we found that the vertical magnetic flux increased instead decreased in the biggest three flares of the AR. After the M6.6 and the X2.2 flares, the unsigned flux increased faster than the pre-flare stage (see Figures 7 a1, c1 and Figure 4 area B). It suggests that there were magnetic emergences in region A and region C, and they were accelerated. Therefore the coronal magnetic collapse is just one reason of the $B_{h}$ enhancement, and the magnetic flux emerged onto the photosphere is another one. The accelerated emergences may link to the flares, but the reason why the emergences of regions A and C are accelerated and the other two do not is not clear yet. Our further work will concentrate on it.

Another difference between regions A, C and regions B, D is in the unsigned helicity. All four regions had a fast and significant increase during the flares and then retrieved to a minimum. However regions A and C had a longer recover time (3-4 h) than regions B and D ($< 2 $ h). It may be due to that regions A and C were bigger and they covered the $\delta$-sunspot of the AR which contains a complex and pivotal magnetic structure. When the bigger, complex and pivotal magnetic structure changed in a major flare, it would take more time to restore.

The decrease caused the decline in the unsigned angular shear of regions A and B during the flares. The unsigned shear of the entire AR had a similar evolution to the unsigned flux, especially in phases II and III. The change of unsigned shear of the four regions did not agree with that of the unsigned flux, even if we enlarge the regions to include the entire polarities. The reason may be that the regions are not independent from the AR in the non-potential aspect or there is an unknown mechanism that makes the regions special.

An interesting correlation is found about the X2.2 flare of the AR. The small area indicated by a box in the $\theta_{as}$ map of Figure 2 has the opposite sign in the angular shear and it is basically cospatial with the source region of the sunquake event which has been investigated by \citet{2011ApJ...734L..15K} and \citet{2011ApJ...741L..35Z}. The area enlarged before the X2.2 flare and shrank after it. Afterwards, a magnetic structure split from a location next to
the area on February 16. It may suggest that this area has different magnetic topology to its vicinity and becomes the key of the outburst event. We will do more research to find the reason of the correlation.

The main results of this work are summarized as follows:

1. The vortex in the source field directly displayed the deflection of $B_{h}$. The deflection was corresponding to the fast rotated sunspots with a time delay, which suggests that the sunspot rotation leads to an increase of the non-potentiality.

2. Two areas that had evident changes of the azimuth of the vector magnetic field were found near the PIL. The change rates of the azimuth were about $1.3^{\circ}$ h$^{-1}$ and $3.6^{\circ}$ h$^{-1}$, respectively.

3. Rapid and prominent increases were found in the variations of unsigned helicity during all four flares in their initial brightening regions. The recovery of the increases took 3-4 h for the two bigger flares, while only took less than 2 h for the other two smaller flares. It may be due to that regions A and C covered the $\delta$-sunspot of the AR which contains a bigger, complex and pivotal magnetic structure for flares.

More AR samples with high quality vector magnetograms are needed to detect the precise role of the non-potentiality.

%
\begin{acknowledgements}
We thank Prof. Jingxiu Wang for his valuable suggestions. The data have been used by courtesy of NASA/SDO and the HMI science team. SDO is a mission for NASA's Living With a Star program. The work is supported by the National Basic Research Program of China under grant 2011CB811403£¬ the National Natural Science Foundation of China (11025315, 10921303, 10973019, 11003024, 40890161, 11203037 and 41074123) and the CAS Project KJCX2-EW-T07.
\end{acknowledgements}

%
%
\bibliographystyle{raa}
\bibliography{Nonpotentiality}

\label{lastpage}

\end{document}